\documentclass{article}
\usepackage{arxiv}
\usepackage[utf8]{inputenc} 
\usepackage[T1]{fontenc}    
\usepackage{lipsum}   
\usepackage{natbib}
\usepackage{doi}

\usepackage{bm}
\usepackage{subfigure}
\usepackage{subcaption}
\usepackage{caption}
\usepackage{amsfonts}
\usepackage{hyperref}
\usepackage{url}
\usepackage{graphicx}
\usepackage{tabularx}
\usepackage{booktabs}
\usepackage{array}
\usepackage{tcolorbox}
\usepackage{xcolor}
\usepackage{tikz}
\usepackage{booktabs}
\usepackage{mdframed}
\usepackage{xcolor}
\usepackage{pifont}
\usepackage{xspace}
\usepackage{multirow}
\usepackage{wrapfig}
\usepackage{bm}
\usepackage{booktabs}       
\usepackage{amsfonts}       
\usepackage{nicefrac}       
\usepackage{microtype}      
\usepackage{xcolor}    
\usepackage{graphicx}
\usepackage{subfigure} 
\usepackage{amsmath}
\usepackage{amssymb}
\usepackage{makeidx}
\usepackage{multicol}
\usepackage{balance}
\usepackage{enumitem}
\usepackage{array}
\usepackage{comment}
\usepackage{adjustbox}
\usepackage{amsthm}
\usepackage{mwe}
\usepackage{cleveref}    

\begin{document}
\title{Confidence Scoring for LLM-Generated SQL in Supply Chain Data
Extraction}
\date{}

\author{
    Jiekal@amazon.com\\
    AWS Supply Chain\\
    \texttt{jiekal@amazon.com}
    \and
    Yikai Zhao\\
    AWS Supply Chain\\
    \texttt{yikai@amazon.com}
}

\maketitle

\begin{abstract}
Large Language Models (LLMs) have recently enabled natural language interfaces that translate user queries into executable SQL, offering a powerful solution for non-technical stakeholders to access structured data. However, one of the limitation that LLMs do not natively express uncertainty makes it difficult to assess the reliability of their generated queries. This paper presents a case study that evaluates multiple approaches to estimate confidence scores for LLM-generated SQL in supply chain data retrieval. We investigated three strategies: (1) translation-based consistency checks; (2) embedding-based semantic similarity between user questions and generated SQL; and (3) self-reported confidence scores directly produced by the LLM. Our findings reveal that LLMs are often overconfident in their own outputs, which limits the effectiveness of self-reported confidence. In contrast, embedding-based similarity methods demonstrate strong discriminative power in identifying inaccurate SQL.

\end{abstract}

\section{Introduction}

Supply chain analysis frequently requires querying large-scale databases to extract insights related to inventory, demand trends, logistics, and supplier performance. Recent advances in Large Language Models (LLMs) have enabled natural language interfaces that allow non-technical users to pose complex questions and receive SQL queries to query databases automatically. While this represents a significant leap in usability and accessibility, it introduces a critical challenge: trust.
In the supply chain context, user queries often involve intricate constraints and domain-specific semantics. For example, a question like \textit{``Which supplier had the fastest average delivery time for Category~X items in the past year?''} requires reasoning and accurate mapping to structured schema elements. LLMs operating in this setting must translate natural language into syntactically valid and semantically precise SQL. These challenges are compounded by the dynamic nature of supply chain databases, where schema variations or out-of-scope queries (e.g., asking for metrics not tracked) are common. Errors in generated SQL or misinterpretation of user intent can result in significant business costs, underscoring the importance of robust reliability mechanisms \cite{li2023large}.
To mitigate this risk, systems must be able to estimate and communicate confidence in their outputs. Confidence estimation enables selective deployment: the model may choose to present results only when sufficiently confident and defer uncertain queries for human verification. This could reduce the likelihood of misinformation.
This paper investigates the novel and practical challenge of estimating confidence for LLM-generated SQL queries in the context of supply chain data extraction. We evaluated three strategies for confidence scoring:
\begin{enumerate}
    \item \textbf{Translation-based consistency checks}, where SQL is re-expressed in natural language and compared to the original question;
    \item \textbf{Embedding-based semantic similarity}, which measures the alignment between the user question and the generated question;
    \item \textbf{Self-reported confidence}, where the LLM assigns its own certainty score during or after generation.
\end{enumerate}

Our goal is to assess the effectiveness of these approaches in identifying unreliable queries and to inform more trustworthy LLM-driven data access systems in supply chain environments.

\section{Methodology}

In this section, we first explain the baseline Text-to-SQL system we used and the evaluation method of the in-distribution text-to-SQL accuracy, and then describe three strategies for estimating confidence in LLM-generated SQL queries, along with relevant supporting literature. We assume an LLM-based text-to-SQL system that takes a user’s natural language question as input and produces an SQL query for a supply chain database. The strategies are applied on top of the base system without altering the underlying LLM answer.

\subsection{Baseline Text-to-SQL System Simulation}
To support accurate SQL generation and evaluation, we first constructed a synthetic question bank by prompting a large language model to generate domain-specific business question templates spanning key supply chain functions such as inventory management, order tracking, demand trends, and supplier performance. Each question is paired with a reference SQL query, accompanying reasoning steps, and verified execution results. All the data is reviewed by domain experts to ensure accuracy. As shown in Table~\ref{table:questionbank}, this high-fidelity dataset serves as a reference for SQL generation and trusted ground truth for evaluating text-to-SQL performance.

\begin{table*}[ht]
  \caption{Example entry from the synthetic question bank used to evaluate SQL generation}
  \label{table:questionbank}
  \begin{tabularx}{\textwidth}{>{\raggedright\arraybackslash}p{3.5cm} X}
    \toprule
    \textbf{Customer Question} & What is the total quantity of \texttt{Apple\_1} ordered? \\
    \midrule
    \textbf{Context} & 
    \texttt{\{ `entities': [\{ `table': `product', `id': `Apple\_1', `description': `Apple' \}, \{ `topic': `inbound\_order\_line' \}] \}} \\
    \midrule
    \textbf{Reasoning} & 
    Based on the input question and context:\newline
    - \texttt{Apple\_1} refers to a product ID in the \texttt{product} table.\newline
    - This ID is used to filter the \texttt{orders} table.\newline
    - The goal is to sum the \texttt{quantity} for this product. \\
    \midrule
    \textbf{SQL (SQLite)} & 
    \texttt{SELECT SUM(quantity) FROM orders WHERE product\_ID = `Apple\_1'} \\
    \midrule
    \textbf{Execution Result} & 
    \texttt{[(40,)]} \\
    \bottomrule
  \end{tabularx}
\end{table*}

To improve SQL generation accuracy, our baseline system adopts a retrieval-augmented generation (RAG) framework, which has been shown to significantly enhance factual correctness and response quality in LLM outputs \citep{lewis2020retrieval, borgeaud2022improving}. Given a customer query, the system first performs entity extraction to identify relevant supply chain concepts (e.g., product ID, Order ID). The These sentence-level embedding model is used to extract entities, the most relevant database schema and a set of five semantically similar example questions from the synthetic question bank. The retrieved schema and question-SQL pairs are provided as contextual prompts to the large language model. This targeted retrieval ensures the model is guided by structurally and semantically aligned examples, promoting syntactic correctness, schema consistency, and domain fidelity in the generated SQL.

\subsection{Evaluation Method}

To evaluate the model's performance and simulate realistic variability, we generated an evaluation dataset by introducing controlled paraphrasing and context variation. Specifically, for each original customer question in the question bank, we used an LLM to generate semantically equivalent rephrasings by altering word choices and entity references ~\cite{hu2022paraphrasing, kumar2020data}. For example, the question \textit{``What's the sales order of Apple\_1 in the last week?''} may yield variants like \textit{``Could you tell me the sales order of Olive\_3 over the last 7 days?''}. This paraphrasing protocol introduces linguistic and contextual diversity while preserving the original intent, thereby testing the model’s generalization to in-distribution examples with increased surface-level complexity ~\cite{dong2020data}. It reflects real-world variability in how business users may phrase similar questions. 
We define in-distribution accuracy as the proportion of evaluation queries for which the LLM-generated SQL query yields the correct result upon execution—i.e., when the output of the generated query matches the ground truth label $Y$. This reflects whether the generated SQL is both syntactically correct and semantically aligned with the input question.
To assess the utility of different confidence scoring strategies, we report the Area Under the Receiver Operating Characteristic Curve (AUROC). AUROC evaluates the model’s ability to discriminate between correct and incorrect SQL predictions based on their associated confidence scores ~\cite{fawcett2006introduction}. Intuitively, it measures the probability that a randomly chosen correct prediction is assigned a higher confidence score than a randomly chosen incorrect one. A perfectly calibrated scoring system yields an AUROC of 1.0, indicating flawless ranking. In contrast, an AUROC of 0.5 implies that the confidence scores are no better than random guessing. 

The AUROC is defined as:

\[
\text{AUROC} = \frac{1}{|\mathcal{P}||\mathcal{N}|} \sum_{x_p \in \mathcal{P}} \sum_{x_n \in \mathcal{N}} I(\cdot)\left(S(x_p) > S(x_n)\right)
\]

$S(x)$ denotes the confidence score assigned to input $x$, and $\mathcal{P}$ and $\mathcal{N}$ represent the sets of positive (correct) and negative (incorrect) examples. $I(\cdot)$ is the indicator function, returning 1 if the argument is true and 0 otherwise. 

\subsection{Confidence Scoring Method}

\paragraph{LLM-Generated Confidence}
Large Language Models (LLMs) can be prompted to articulate or quantify their confidence, either directly within a single response or through a follow-up query. The self-probing approach was introduced first by \citet{kadavath2022language}. After generating an answer, the model is asked a binary question such as ``Is the previous answer correct?'' The predicted probability of a positive response serves as a proxy of confidence. This method demonstrated up to 30\% improvement in identifying incorrect answers compared to relying solely on raw token probabilities. Subsequent studies have built on the concept showing that prompting the model to first explain its reasoning and then report a numerical confidence score led to better-calibrated “predictions” ( \citet{tian2023justifications}). The experiments indicated that the model's self-reported confidence was more closely aligned with actual correctness than its conditional likelihood, outperforming log-probability-based by approximately 50\% in calibration metrics across question-answering tasks. These findings underscore the potential of prompting strategies for eliciting reliable confidence estimates from LLMs.

\paragraph{Translation-based Consistency Check}
This approach leverages the intuition that if the SQL truly captures the user's intent, its natural language rendition should semantically match the input question. A self-validation framework called CycleSQL was presented to implement the translation-based consistency check idea \citet{Fan2024}. In this approach, after an LLM generates a SQL query, the system executes the query and translates the results back into natural language, essentially producing an answer explanation. It then uses a textual entailment model to compare the natural language explanation with the original question, treating it as a premise-hypothesis pair for natural language inference. If the explanation does not entail the question (indicating a discrepancy in semantics), the SQL is flagged as wrong and the model attempts to refine it. This round-trip verification process significantly boosts accuracy on benchmarks, as the model catches subtle errors where the SQL, though syntactically valid, fails to answer the questions.

\paragraph{Embedding Similarity Approach}

Another approach is using embeddings to find if a new input question is semantically similar to some known questions with verified answers. If a closely similar question has appeared before in the question bank with verified answer, one can infer that the LLM’s answer to the new question is more likely correct. Conversely, if the question is unlike anything seen before, the model’s output should be treated with lower confidence. This technique has been observed in real-world deployments of QA systems to improve consistency. For instance, an LLM-augmented search application might remember that it correctly answered \textit{``How many passengers fit in a Boeing 747?} before, so when a semantically equivalent question \textit{``What is the seating capacity of a 747 jumbo jet?} is asked, the model’s answer can be cross-checked against the previous answer for agreement. If they align, the confidence in the answer increases; if not, it may indicate uncertainty. Such semantic caching exploits question-to-question similarity to avoid repeat mistakes and to provide calibrated confidence based on prior successes.

\section{Experiments}
In this section, we present the experimental setup and confidence score accuracy results obtained. The experiments were designed to evaluate the effectiveness of our proposed confidence estimation strategies within a controlled, privacy-preserving environment. These strategies were integrated into our base SQL generation system to assess their applicability and performance.

\subsection{Experimental Setup}
To ensure the customer data privacy, we demonstrated our method evaluation using a synthetic dataset that we generated. The synthetic dataset consisting of 988 supply chain related questions ~\cite{trivedi2023llmsupplychain}. There are 731 simpler questions and 257 complex business questions. Simpler questions typically map to SQL queries with a well-defined and consistent structure. For example, the question \textit{``What’s the total quantity of item X sold yesterday?''} corresponds to a straightforward aggregation query with minimal variation. In contrast, business questions often involve domain-specific terminology and admit multiple valid SQL formulations. Consider the question \textit{``How many sites have stock-out risk for item Y?''} The term \textit{``stock-out risk''} is a supply chain concept with a fixed SQL definition, typically derived from business logic. Such queries increase semantic ambiguity and structural variability, posing greater challenges for SQL generation systems. All experiments were conducted using Claude Sonnet 3, the most advanced model available at the time of experimentation, accessed through Amazon Bedrock API ~\cite{amazonbedrock2023} with default hyperparameter settings. The same model was used for both SQL generation and downstream confidence estimation tasks. Unless otherwise stated, no fine-tuning or task-specific training was performed.

Our evaluation comprises three stages, each exploring a different approach to confidence score generation for LLM-produced SQL queries.

\paragraph{Stage 1: Prompt-Based Self-Reported Confidence}
In the first stage, we evaluated the ability of the LLM to self-report its confidence in the generated SQL query. We experimented with multiple prompting strategies to elicit confidence scores in binary (e.g., True or False) and scalar (e.g., score 0-100) formats. Details of the prompt templates used are provided in Table~\ref{table:prompt-variants}. This setup assesses whether the model’s self-assessed confidence aligns with actual correctness.

\begin{table}[ht]
  \caption{Prompt strategies for eliciting self-reported confidence scores from the LLM}
  \label{table:prompt-variants}
  \small
  \begin{tabularx}{\columnwidth}{>{\raggedright\arraybackslash}p{3cm} X}
    \toprule
    \textbf{Prompt Strategy} & \textbf{Prompt Template} \\
    \midrule
    Subtracting from 100 & Aim for a wide distribution of scores that accurately reflects the varying levels of confidence you have in the generated SQL. \\
    
    Money Betting & Aim to place bets that accurately reflect your true confidence in the generated SQL, as if real money were at stake. \\
    
    SQL Component Weighting & Calculate the overall confidence score by taking the weighted average of the component scores, with weights determined by the importance of each component in the given context. \\
    
    Straight Instructions & After generating the SQL query, please take a moment to critically analyze your output and generate your own confidence score, between 0 and 100, about the generated SQL. \\
    
    Binary Classification & Generate the binary judgment of “confident” or “not confident”; explain why you gave that classification within the \texttt{<conf\_reason>...</conf\_reason>} XML tags. \\
    \bottomrule
  \end{tabularx}
\end{table}

\paragraph{Stage 2: Translation-Based Consistency Scoring}
In the second stage, we implemented the translation-based validation framework. For each LLM-generated SQL query, we prompt a second LLM to generate a natural language question that the SQL is likely to answer, given the SQL statement, reasoning trace, and execution result. The inferred question is then compared to the original input question using a third LLM acting as a natural language judge. If the inferred question is semantically equivalent to the original, the SQL is considered high-confidence. 

\paragraph{Stage 3: Embedding-Based Similarity Scoring}
In the third stage, we adopted an embedding-based approach to estimate confidence through semantic alignment. Specifically, we compute the cosine similarity between the input customer question and the five semantically relevant examples retrieved from the question bank for in-context learning. The average similarity score across these five examples is used as the confidence estimate for the generated SQL query. This method assumes that if the target question closely resembles high-quality examples, the model is more likely to produce reliable SQL.

\subsection{Results}
\paragraph{Stage 1}
As shown in Table~\ref{table:prompt-auroc}, the binary classification prompt strategy achieved the highest AUROC among all evaluated methods, outperforming direct scalar score generation approaches. However, the overall performance across all strategies remains close to the 0.5 threshold, indicating limited discriminative power. While analyzing the mismatch between confidence scores and actual accuracy, we observed that the LLM tends to assign consistently high confidence scores—often above 0.9—to its own generated SQL queries, particularly for simpler questions. This suggests that the model’s self-reported confidence scores are poorly calibrated and often overconfident—offering little improvement over random guessing in identifying correct versus incorrect SQL outputs.
\begin{table}[ht]
  \caption{AUROC performance of different prompt strategies for self-reported confidence estimation in stage 1}
  \label{table:prompt-auroc}
  \small
  \begin{tabularx}{\columnwidth}{>{\raggedright\arraybackslash}X r}
    \toprule
    \textbf{Prompt Strategy} & \textbf{AUROC} \\
    \midrule
    Subtracting from 100 & 0.544 \\
    Money Betting & 0.509 \\
    SQL Component Weighting & 0.481 \\
    Straight Instructions & 0.523 \\
    Binary Classification & \textbf{0.553} \\
    \bottomrule
  \end{tabularx}
\end{table}

Table ~\ref{table:prompt-auroc} shows the classification prompt strategy will provide better AUROC than  direct score generation. 
Overall, the results show that it's over-confident for this method. All of the AUROC are near 0.5 borderline, meaning the confidence score is not much better than a random guess. 
\paragraph{Stage 2}
Table ~\ref{table:confusion-matrix} presents the distribution of confident versus non-confident predictions across accurate and inaccurate SQL generations. The binary confidence classifier achieved an AUROC of 0.524, indicating performance only marginally better than random chance. These results suggest that the strategy remains overconfident and its confidence scores fail to reliably distinguish between correct and incorrect SQL outputs. The translation-based approach demonstrated lower performance on complex questions, particularly due to hallucinations around domain-specific terminology. For instance, the LLM often translates the term "quantity" as "volume" during SQL interpretation. While both terms can imply a numerical measure in general language, they carry distinct meanings in the supply chain domain: quantity typically refers to the number of units ordered, whereas volume refers to the physical size or space occupied by the items. This kind of domain-specific ambiguity in natural language is difficult for the model to detect and disambiguate, often resulting in overconfident yet incorrect SQL generation.

\begin{table}[ht]
  \caption{Confusion matrix of binary confidence classification vs. actual SQL correctness in stage 2}
  \label{table:confusion-matrix}
  \small
  \centering
  \begin{tabular}{lcc}
    \toprule
    \textbf{Confidence Accuracy} & \textbf{Confident} & \textbf{Non-Confident} \\
    \midrule
    Accurate     & 754 & 12 \\
    Non-Accurate & 208 & 14 \\
    \bottomrule
  \end{tabular}
\end{table}
\paragraph{Stage 3}

\begin{figure}[htb!]
    \centering
    \includegraphics[width=\linewidth]{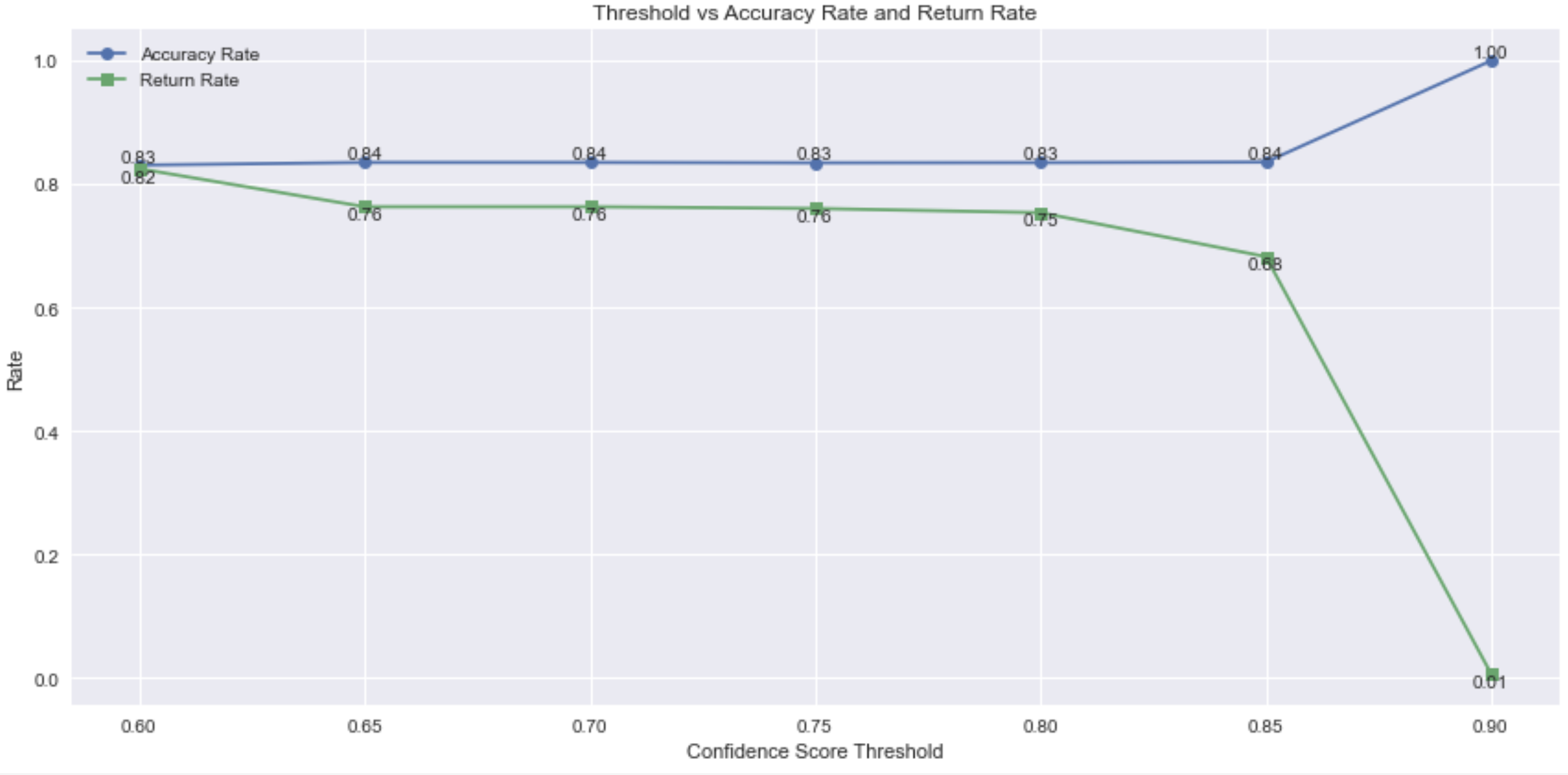}
    \caption{Simple Questions: Accuracy and Returning Rate by Confidence Level}
    \label{fig:simple}
\end{figure}

\begin{figure}[htb!]
    \centering
    \includegraphics[width=\linewidth]{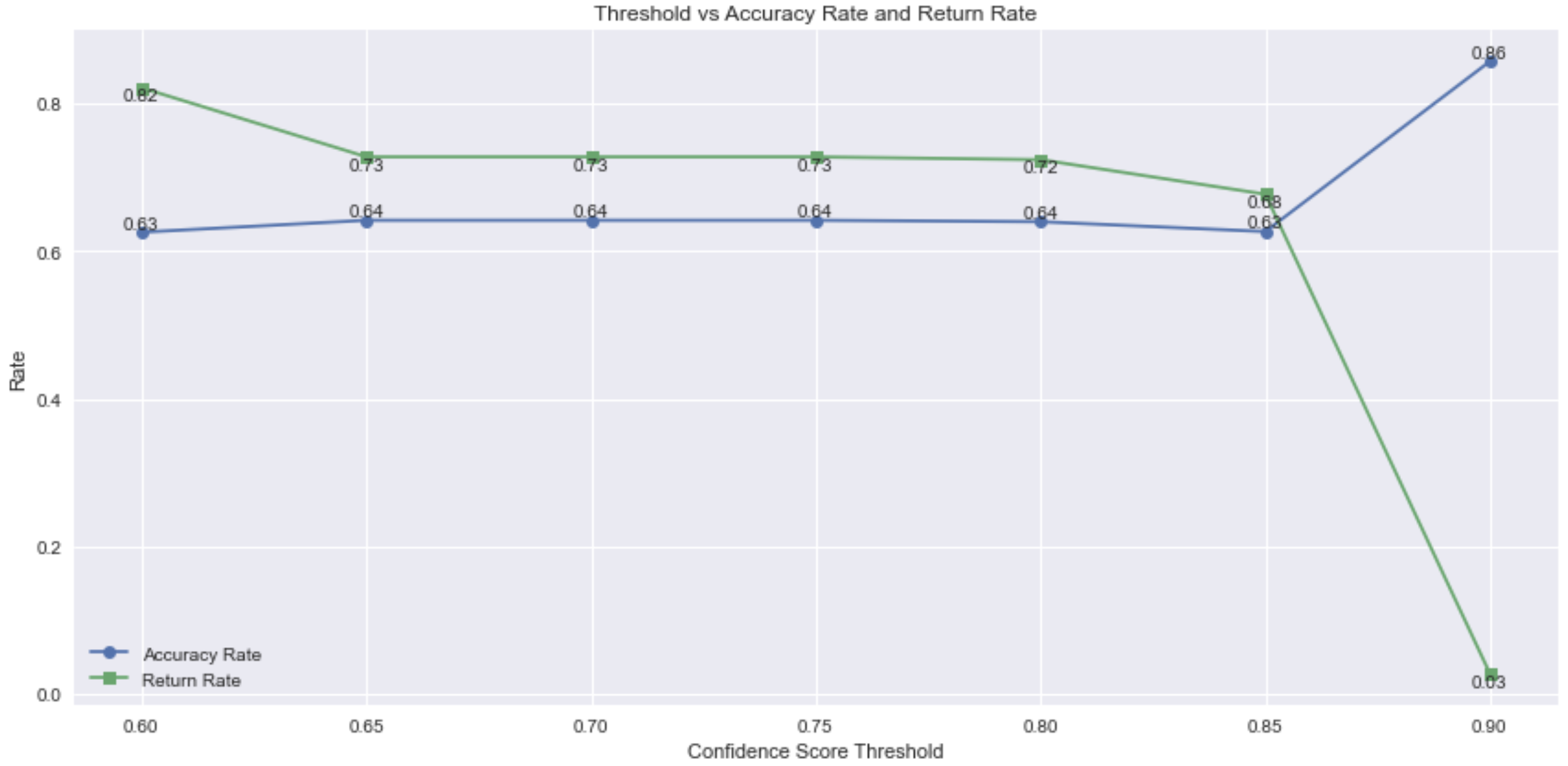}
    \caption{Complex Questions: Accuracy and Returning Rate by Confidence Level}
    \label{fig:business}
\end{figure}

The similarity-based confidence score achieved an AUROC of 0.57 across both simple questions and complex questions, the highest among all methods evaluated. Figure~\ref{fig:simple} illustrates the relationship between SQL execution accuracy, return rate (i.e., the proportion of queries with confidence above a given threshold), and average confidence score for simple questions. Figure~\ref{fig:business} presents the same analysis for complex business questions.

Both figures reveal a clear pattern: when the confidence score exceeds a specific threshold, empirically observed around 0.85, the majority of generated SQL queries attain significantly higher execution accuracy. This indicates that the similarity score serves as a reliable proxy for SQL correctness and can effectively be used to filter out low-confidence outputs, thereby improving the overall trustworthiness of the system.

\section{Conclusion}

In this study, we evaluated three distinct approaches for estimating confidence levels in LLM-generated SQL queries: (1) self-reported scores via direct prompting, (2) translation-based consistency checks, and (3) embedding-based similarity scoring.

Our experimental results reveal a consistent trend: large language models tend to be overconfident in the SQL generation to extract supply chain data, regardless of the prompting strategy employed. Among the self-assessment methods, the binary classification prompt outperformed scalar score generation, though still with limited discriminative ability. In contrast, the embedding-based similarity score demonstrated the stronger correlation with SQL correctness, achieving the higher AUROC and showing clear utility in separating high-quality from low-quality generations.

As a practical insight for real-world deployment, we find that the similarity score offers a promising mechanism for thresholding LLM outputs—particularly in complex supply chain data extraction scenarios, where schema diversity and business logic can increase error risk. By filtering out low-confidence queries based on semantic alignment, systems can significantly improve overall reliability and user trust.

Future work will explore the generalizability of the similarity-based confidence strategy across diverse supply chain datasets, schemas, and query intents. In addition, integrating confidence-driven post-filtering into end-to-end decision support systems may offer further gains in production robustness.

\section{Limitation and Future Work}
One limitation of this study is that all experiments were conducted on a synthetic dataset due to customer data privacy concerns. While the synthetic data allowed us to prototype and evaluate the confidence scoring approach in a controlled environment, it may not fully capture the complexities of real-world supply chain data. In future work, we aim to extend our evaluation to public supply chain datasets, such as the M5 competition dataset, to validate the generalizability and robustness of our approach under more realistic conditions.
Additionally, Large Language Models are evolving rapidly. Although this study used Claude Sonnet 3, the latest model available to us at the time, it is important to evaluate the proposed confidence scoring strategy across a broader set of LLMs. Future work will explore the performance of the method on newer and more diverse models, including GPT-4, DeepSeek, and Claude Sonnet 3.5 or 3.7. As different models, particularly those optimized for reasoning, may demonstrate varying levels of confidence reliability, such comparison is essential to assess the method’s transferability and performance consistency.
Lastly, our current confidence scoring framework is built on a baseline text-to-SQL system that leverages a retrieval-augmented generation pipeline. As more domain-specific supply chain databases are developed, incorporating fine-tuning strategies on top of this foundation may further improve both SQL generation accuracy and the reliability of the associated confidence scores. Future exploration into fine-tuning will allow for deeper model alignment with domain-specific language and schema patterns, thereby enhancing practical utility in enterprise settings.
\newpage
\bibliographystyle{unsrtnat}
\bibliography{references} 

\end{document}